\documentclass[aps,amsfonts,amsmath,prd,preprint,nofootinbib]{revtex4-1}
\newcommand{\beq}{\begin{equation}}
\newcommand{\eeq}{\end{equation}}

\def\be{\begin{equation}}
\def\ee{\end{equation}}
\def\beqa{\begin{eqnarray}}
\def\eeqa{\end{eqnarray}}

\usepackage{epsfig}

\begin{document}

\title{Measures for a  Transdimensional Multiverse}

\author{Delia Schwartz-Perlov}
\author{Alexander Vilenkin}

\affiliation{Institute of Cosmology, Department of Physics and Astronomy,\\ 
Tufts University, Medford, MA 02155, USA}

\def\changenote#1{\footnote{\bf #1}}

\begin{abstract}

The multiverse/landscape paradigm that has emerged from eternal
inflation and string theory, describes a large-scale multiverse
populated by ``pocket universes'' which come in a huge variety of
different types, including different dimensionalities.  In order to
make predictions in the multiverse, we need a probability measure.  In
$(3+1)d$ landscapes, the scale factor cutoff measure has been
previously shown to have a number of attractive properties.  Here we
consider possible generalizations of this measure to a
transdimensional multiverse.  We find that a straightforward extension
of scale factor cutoff to the transdimensional case gives a measure
that strongly disfavors large amounts of slow-roll inflation and
predicts low values for the density parameter $\Omega$, in conflict
with observations.  A suitable generalization, which retains all the
good properties of the original measure, is the ``volume factor''
cutoff, which regularizes the infinite spacetime volume using cutoff
surfaces of constant volume expansion factor.

\end{abstract}

\maketitle
\section{Introduction}

Eternal inflation and string theory both suggest that we live in a
multiverse which is home to many different types of vacua
\cite{Vilenkin:1983xq, Linde:1986fd, Bousso:2000xa, KKLT, Susskind}. 
Many of these vacua are high-energy metastable states that can decay
to lower-energy states through bubbles which nucleate and expand
in the high-energy vacuum background \cite{CdL}. 
Inverse transitions are also possible: 
bubbles of high-energy vacuum can nucleate in low-energy ones 
\cite{EWeinberg}\footnote{This is the so-called
  ``recycling'' process. In 
both cases, the radius of the bubbles asymptotically approaches the 
comoving horizon size in the parent vacuum at the moment of nucleation
\cite{recycling}.}.  
Once the process of eternal inflation begins,  an unbounded number of
bubbles will be created, one within the other, of every possible type.  
In addition each bubble has infinite spacelike slices when described
in the maximally symmetric FRW coordinate system, and thus infinite
spatial volume.  
The multiverse spacetime therefore has an
infinite number of observers of every possible type. 

In order to make predictions, we need to be able to
compare the numbers of different types of events that can occur in the
multiverse.   This is non-trivial because of the aforementioned
infinities, and has become known as ``the measure problem"
\cite{MPreviews}. 
A number of measure proposals have been developed
~\cite{LM,LLM,GBLL,AV94,AV95,Linde06,Linde07,GTV,pockets,GSPVW,ELM,
diamond,censor,Vanchurin07,Winitzki08,LVW,DeSimone:2008bq,BB2,Guthlec}, 
and some have already been ruled out, as they either lead to paradoxes
or yield predictions in conflict with observations. All these measures
have been defined within the context of regular ``equidimensional"
$(3+1)d$ eternal inflation: all parent and daughter vacua have the
same dimensionality.  However, in general we expect the string theory
landscape to allow parent vacua to nucleate daughter vacua with
different effective dimensionality \cite{Zelnikov}.  The purpose of
the present work is to introduce a measure for a ``transdimensional''
multiverse.

One of the most attractive $(3+1)d$ measure proposals is the scale
factor cutoff measure or, for brevity, scale factor measure
\cite{LM,LLM,GBLL,AV94,AV95}. It does not suffer from any
obvious problems and gives prediction for the cosmological constant in
a good agreement with the data \cite{DeSimone:2008bq,BB2,Guthlec}. 
In this measure the infinite spacetime volume of the multiverse is
regularized with a cutoff at surfaces of constant scale factor.  More
precisely, one chooses some initial spacelike hypersurface $\Sigma_0$,
where the scale factor is set to be equal to one, $a_0 = 1$, 
and follows the congruence of timelike geodesics orthogonal to that
surface until the scale factor reaches some specified value $a_{max}$.
In the $(3+1)$-dimensional case, the scale factor is defined as 
\beq
a ={\cal V}^{1/3}, 
\label{aV}
\eeq
where ${\cal V}$ is the volume expansion factor along
the geodesics.  Relative probabilities are calculated by counting the
number of times a given event occurs in the finite spacetime volume
spanned by the congruence, between the initial ($\Sigma_0$) and final
($\Sigma_{max}:a=a_{max}$) hypersurfaces, in the limit that the scale
factor cutoff $a_{max} \rightarrow \infty$.

In this paper we consider possible extensions of the scale factor
measure to a transdimensional landscape.  Eq.~(\ref{aV}) defining the
scale factor can be generalized as
\beq
a ={\cal V}^{1/D}, 
\label{aVD}
\eeq 
where $D$ is the number of expanding (non-compact) dimensions.  As  a
straightforward extension of the $(3+1)d$ scale factor measure, one
can still use constant-$a$ surfaces as a cutoff.  We will find, however,
that this prescription leads to a problem: it strongly disfavors large
amounts of slow-roll inflation inside bubbles and predicts low values
of the density parameter $\Omega$, in conflict with observations.

An alternative prescription is to use constant volume factor surfaces,
${\cal V}=const$, for a cutoff.  We shall call this the volume factor
measure.  In an equidimensional multiverse, a surface of constant
scale factor is also a surface of constant volume factor.  However,
for transdimensional universes, different vacua generally have
different values of $D$, so constant scale factor and volume factor
surfaces are not the same.  We will find that the volume factor cutoff
results in an expression for comparing relative probabilities that is
analogous to that of the equidimensional scale factor measure and
provides a natural framework in which to extend the successful
features of the scale factor measure to transdimensional landscapes.  

The rest of the paper is organized as follows: In Section
\ref{regularscalefactor} we review the scale factor measure and
outline how it can be implemented using the rate (master) equation
formalism in four spacetime dimensions.  In Section III we review the
spacetime structure of a transdimensional multiverse using the
6-dimensional Einstein-Maxwell theory as an example and generalize the
master equation to the transdimensional case.  This formalism is then
applied in Section IV to investigate the properties of the
transdimensional volume factor and scale factor measures.  Our
conclusions are briefly summarized and discussed in Section V.

\section{The scale factor measure in $(3+1)d$} \label{regularscalefactor} 

In this section we shall review the key features of the scale factor
measure as found in Refs.~\cite{DeSimone:2008bq,BB2,Guthlec}.   

The scale factor time $\eta$ is defined in terms of the expansion
factor of the geodesics $a(\tau)$ as  
\beq
\eta=\ln{a}
\eeq
and is related to the proper time $\tau$ via 
\beq
d \eta = H(\tau) d\tau.
\eeq
Here, $H(\tau)$ is the local expansion rate of the geodesic congruence,
which can be defined as the divergence of the four-velocity field
$u^{\mu}(x)$ of the congruence,   
\beq
H= (1/3)u^{\mu}_{~;\mu}.
\label{H}
\eeq

We start with a congruence of geodesics orthogonal to the initial
spacelike hypersurface $\Sigma_0$ and allow them to expand until the
local scale factor reaches a critical value $a_{max}$ corresponding to
$\eta_{max}$.   This defines the cutoff hypersurface $\Sigma_{max}$.   
The relative probability to make an observation of type $i$ or $j$ is
given by the ratio of the number of times that each type of
observation ($\mathcal{N}_i$ or $\mathcal{N}_j$) occurs in the
spacetime volume spanned by the congruence,  
in the limit that $\eta_{max} \rightarrow \infty$, 
\beq
{{P(O_i)} \over {P(O_j)}} \equiv \lim_{\eta_{max} \rightarrow \infty }
{{\mathcal{N}_i}\over{\mathcal{N}_j}} \label{ninjpipj} 
\eeq 

An intuitive picture of the scale factor cutoff emerges when we think
of the geodesic congruence as a sprinkling of test ``dust" particles.
Imagine that this dust has a uniform density $\rho_0$ on our initial
hypersurface $\Sigma_0$.  As the universe expands, the dust density
(in the dust rest frame) dilutes.   The scale factor cutoff is then
initiated when the density drops below a critical value. The scale
factor time can be expressed in terms of the dust density as  
\beq
\eta = - 1/3 \ln{{{\rho}\over{\rho_0}}}. 
\label{etarho}
\eeq
We can also introduce the volume expansion factor ${\cal
  V}=\rho_0/\rho$ and the corresponding ``volume time''
\beq
T=\ln {\cal V}.
\eeq
In a $(3+1)d$ multiverse this time is trivially related to the scale
factor time,
\beq
T=3\eta.
\eeq

The definition of the scale factor time $\eta$ in terms of a geodesic
congruence encounters ambiguities in regions of structure formation,
where geodesics decouple from the Hubble flow and eventually
cross.  In particular, Eq.~(\ref{etarho}) suggests that scale factor
time comes to a halt in bound structures, like galaxies, while it
continues to flow in the intergalactic space.  It is not clear whether
one ought to use this local definition of $\eta$ for the scale factor
cutoff, or use some nonlocal time variable that is more closely
related to the FRW scale factor.  Possible ways of introducing such a
variable have been discussed in Refs.~\cite{BB2,BB1}.  Here we shall
assume that some such nonlocal procedure has been adopted, so that
structure formation has little effect on the cutoff hypersurfaces.

The numbers of observations ${\cal N}_{i,j}$ in Eq.~(\ref{ninjpipj})
are proportional to appropriately regularized volumes occupied by the
corresponding vacua.  We shall now review how these volumes can be
calculated using the master equation of eternal inflation.

\subsection{The rate equation}

The comoving volume fractions $f_i(\eta)$  occupied by vacua of type
$i$ on a constant scale factor time slice $\eta$ obeys the rate
equation (also called the master equation)
\cite{recycling,GSPVW} 
\beq
{df_i\over{d\eta}}=\sum_j (-\kappa_{ji}f_i + \kappa_{ij}f_j).
\label{dfdt}
\eeq
Here, the first term on the right-hand side accounts for loss of
comoving volume due to bubbles of type $j$ nucleating within those of
type $i$, and the second term reflects the increase of comoving volume
due to nucleation of type-$i$ bubbles within type-$j$ bubbles.

The transition rate $\kappa_{ij}$ is defined as the probability per
unit scale factor time for a geodesic "observer" who is currently in vacuum $j$ to find
itself in vacuum $i$, 
\beq
\kappa_{ij}=\Gamma_{ij}{4\pi\over{3}}H_j^{-4},
\label{kappa}
\eeq
where $\Gamma_{ij}$ is the bubble nucleation rate per unit
physical spacetime volume,
\be
H_j = (\Lambda_j/3)^{1/2}
\label{Hj}
\ee
is the expansion rate and $\Lambda_j$ is the energy density in vacuum $j$.
The total transition rate out of vacuum $i$ is given by the sum
\beq
\kappa_{i}=\sum_j \kappa_{ji}.
\label{kappai}
\eeq

We distinguish between the recyclable, non-terminal vacua with
$\Lambda_j>0$, which are sites of eternal inflation and further bubble
nucleation, and the non-recyclable, ``terminal vacua", for which
$\Lambda_j\leq 0$.  Then, by definition, $\Gamma_{ij}=0$ when the
index $j$ corresponds to a terminal vacuum.

Eq.~(\ref{dfdt}) can be written in a vector form,
\beq
{d{\bf f}\over{d\eta}}={\mathbf M}{\bf f},
\label{matrixf}
\eeq
where ${\bf f(\eta)}\equiv \{ f_j(\eta)\}$ and
\beq
M_{ij}=\kappa_{ij}-\delta_{ij}\sum_r \kappa_{ri}.
\label{Mij}
\eeq
The asymptotic solution of (\ref{matrixf}) at large $\eta$ has the
form
\beq
{\bf f}(\eta)={\bf f^{(0)}}+{\bf s}e^{-q \eta}+ ...
\label{asympt}
\eeq
Here, ${\bf f}^{(0)}$ is a constant vector
which has nonzero components only in terminal vacua.  Any such
vector is an eigenvector of the matrix ${\mathbf M}$ with zero
eigenvalue, 
\beq {\mathbf M}{\bf f_0}=0. 
\eeq 
As shown in
\cite{GSPVW}, all other eigenvalues of $\mathbf{M}$ have a negative
real part, so the solution approaches a constant at late times.
Moreover, the eigenvalue with the smallest (by magnitude) real part is
purely real.  We 
have denoted this eigenvalue by $-q$ and the corresponding eigenvector by
$\mathbf{s}$.

The asymptotic values of the
terminal components $f_j^{(0)}$ depend on the choice of initial
conditions. For any physical choice, we should have $f_j^{(0)} \geq
0$. It can also be shown \cite{GSPVW} that $s_j
\leq 0$ for terminal vacua and $s_j \geq 0$ for recycling vacua.
In the case of recycling vacua, it follows from Eq.~(\ref{asympt}) that
\beq
f_j \approx s_j e^{-q \eta}.
\label{fs}
\eeq

The physical properties of bubble interiors, such as
the spectrum of density perturbations, the density parameter, etc.,
depend not only on the vacuum inside the bubble, but also on the
parent vacuum that it tunneled from (or, more precisely, on the
properties of the bubble implementing the transition between the two
vacua).  Bubbles should therefore be 
labeled by two indices, $ij$, referring to the daughter and parent
vacua respectively.  The fraction of comoving volume in type-$ij$
bubbles, $f_{ij}$, satisfies a slightly modified rate equation
\cite{Aguirre} 
\beq
{df_{ij}\over{d\eta}} = \sum_k (-\kappa_{ki}f_{ij} +
\kappa_{ij}f_{jk})  = -\kappa_i f_{ij} + \kappa_{ij}f_j .
\label{dfijdt}
\eeq
Here and below we do not assume summation over repeated indices,
unless the summation is indicated explicitely.
The fractions $f_{ij}$ and $f_i$ are related by 
\beq
f_i = \sum_j f_{ij} ,
\label{fifij}
\eeq
and the original rate equation (\ref{dfdt}) can be recovered from
(\ref{dfijdt}) by summation over $j$.  Solutions of Eq.~(\ref{dfijdt})
can also be expressed in the form (\ref{asympt}), and for non-terminal
bubbles we have
\beq
f_{ij} \approx s_{ij} e^{-q \eta} .
\eeq
Note that it follows from (\ref{fifij}) that the eigenvalue $q$ for
the modified equation (\ref{dfijdt}) should be the same as for the
original equation.

The comoving volume fractions found on equal scale factor time
hypersurfaces are directly related to the physical volume fractions.
Thus we find that asymptotically the physical volumes in nonterminal
bubbles are given by 
\beq
V_{ij}(\eta)=V_0s_{ij} e^{(3-q)\eta} 
\label{asymphysvol},
\eeq
where $V_0$ is the volume of the initial hypersurface $\Sigma_0$ and
$e^{3\eta}$ is the volume expansion factor. 
Note that the growth of volume is given by $e^{\gamma \eta}$, where
\beq
\gamma=3-q
\eeq
is the fractal dimension of the inflating region \cite{Aryal}.  
This is slightly less than $3$, because volume is lost to terminal vacua. 

Assuming that upward transitions in the landscape are strongly
suppressed, the eigenvalue $q$ can be approximated as
\cite{SchwartzPerlov:2006hi} $q\approx \kappa_{min}$, 
where $\kappa_{min}$ is the decay rate of the slowest-decaying vacuum.
In general, it can be shown \cite{BB1} that 
\beq
q \leq \kappa_{min}.
\label{qkappamin}
\eeq  
In the string theory landscape, $\kappa_{min}$ is expected to be
extremely small, and thus $\gamma$ is very close to 3. 

The volume fractions can also be expressed in terms of the volume time
$T$, which in our case is simply $T=3\eta$,
\beq
V_{ij}(T)=V_0s_{ij} e^{\tilde{\gamma} T} 
\label{voltasymphysvol},
\eeq
where we have defined
\beq
{\tilde\gamma}=1-{\tilde q}
\label{tildegamma}
\eeq
and
\beq
\tilde{q} = q/3.
\label{tildeq}
\eeq

\subsection{The number of observers}

Bubble interiors can be described by open FRW coordinates
$(\tau,\xi,\theta,\phi)$, 
\beq
ds^2 = -d\tau^2 + a_{ij}^2(\tau)(d\xi^2+\sinh^2\xi d\Omega^2) ,
\label{bubbleFRW}
\eeq
where the functional form of $a_{ij}(\tau)$ is generally different in
different bubbles.  We shall assume for simplicity that all
observers in bubbles of type $ij$ are formed at a fixed 
FRW time $\tau^{obs}_{ij}$ with some average density $n_{ij}^{obs}$.
This defines the ``observation hypersurface'' within each habitable
bubble.  The infinite three-volume of this hypersurface is regularized
by the scale factor cutoff at some $\eta = 
\eta_{max}$.  The regulated number of type-${ij}$ observers is 
\beq
\mathcal{N}_{ij}^{obs}= V_{ij}^{obs} n_{ij}^{obs} ,
\label{normaldecompreg}
\eeq
where $V_{ij}^{obs}$ is the total volume of observation hypersurfaces
in all type-$ij$ bubbles. 

For our purposes in this paper, it will not be necessary to
distinguish between the number of observers and the number of
observations, so we shall use these terms interchangeably.  If one is
interested in some particular kind of observation, then $n_{ij}^{obs}$ in
Eq.~(\ref{normaldecompreg}) should be understood as the density of
observers who perform that kind of observation.

The calculation of the volume $V_{ij}^{obs}$ has been discussed in
Refs.~ \cite{Vilenkin:1996ar, BB2, Guthlec, Bousso:2007nd}.   
Here we give a simplified derivation, referring the reader to the
original literature for a more careful analysis.   

As the geodesics cross from the parent vacuum into the bubble, they
become comoving with respect to the bubble FRW coordinates
(\ref{bubbleFRW}) within a few Hubble times.  This implies that a
fixed amount of FRW time $\tau$ roughly  corresponds to a fixed amount
of scale factor time $\eta$ along the geodesics.  (This will be
increasingly accurate at large $\tau \gg H^{-1}$.)  We shall denote
$\Delta\eta_{ij}^{obs}$ to be the amount of scale factor time between
crossing into a bubble and reaching the observation hypersurface
$\tau=\tau_{ij}^{obs}$.  If the expansion rate in the daughter bubble right
after nucleation is not much different from the expansion rate $H_j$
in the parent vacuum, this can be approximated as 
\beq
\Delta \eta_{ij}^{obs} \approx \ln[H_j a_{ij}(\tau_{ij}^{obs})] .
\label{Deltaeta}
\eeq

From Eq.~(\ref{dfijdt}), the comoving volume fraction that crosses
into $ij$ bubbles per time interval $d\eta$ is 
\beq
d^+ f_{ij} = \kappa_{ij}f_j (\eta) d\eta. 
\eeq
The corresponding physical volume is
\beq
d^+ V_{ij} = V_0 e^{3\eta}d^+ f_{ij} = V_0 \kappa_{ij} s_j e^{(3-q)\eta} d\eta ,
\eeq
where we have used Eq.~(\ref{fs}) for the comoving volume fraction in
the recycling parent vacuum $j$.  The total regulated volume of
observation hypersurfaces 
in type-$ij$ bubbles can now be expressed as 
\beq
V_{ij}^{obs} = V_0 e^{3\Delta\eta_{ij}^{obs}}
e^{-\lambda_{ij}\Delta\eta_{ij}^{obs}}
\int_0^{\eta_{max}-\Delta\eta_{ij}^{obs}} \kappa_{ij} s_j
e^{(3-q)\eta} d\eta . 
\label{Vijintegral}
\eeq
The factor $\exp(3\Delta\eta_{ij}^{obs})$ in front of the integral
accounts for volume expansion in bubble interiors, and the factor
$\exp(-\lambda_{ij} 
\Delta\eta_{ij}^{obs})$ accounts for the volume loss due to bubble nucleation in
the course of slow-roll inflation and subsequent non-vacuum evolution
inside the bubbles.  The corresponding nucleation rate is generally different
from that in the vacuum; that is why we used a different notation.
The integration in (\ref{Vijintegral}) is up to $\eta_{max}
-\Delta\eta_{ij}^{obs}$: regions added at later times are cut 
off before they evolve any observers.   Performing the integration, we find
\beq
V_{ij}^{obs} = {1\over{3-q}}V_0 \kappa_{ij}s_{j} e^{(3-q)\eta_{max}}
e^{(q-\lambda_{ij})\Delta\eta_{ij}^{obs}} .
\label{Vijresult}
\eeq

The constant ($ij$-independent) factors in
Eq.~(\ref{Vijresult}) do not affect the relative numbers of
observers in different vacua.  Dropping these irrelevant factors and
using Eq.~(\ref{normaldecompreg}), we obtain 
\beq
\mathcal{N}_{ij}^{obs}\propto \kappa_{ij}s_j n_{ij}^{obs}
e^{(q-\lambda_{ij})\Delta\eta_{ij}^{obs}} . 
\label{Nij}
\eeq

The decay rate $\lambda_{ij}$ is typically exponentially
suppressed, and according to Eq.~(\ref{qkappamin}), the eigenvalue $q$
should also be extremely small.  One can expect therefore that, to a
good approximation, the last factor in (\ref{Nij}) can be dropped.



\section{A transdimensional multiverse} 

\subsection{Einstein-Maxwell landscape}

We now wish to generalize the above formalism to a transdimensional
multiverse.  We will try to keep the discussion general, but it will
be helpful to have a specific model in mind.  So we start by reviewing
the $6d$ Einstein-Maxwell model, which has been used as a toy model
for flux compactification in string theory \cite{Douglas} and has been
recently studied in the multiverse context in
Refs. \cite{BlancoPillado:2009di,Carroll:2009dn,BlancoPillado:2009mi}.  

The model includes Einstein's gravity in $(5+1)$ dimensions, a Maxwell
field, and a positive cosmological constant. It admits solutions in
which some of 
the spatial dimensions are compactified on a sphere.  These extra
dimensions are stabilized by either magnetic or electric flux.  The
landscape of this theory has turned out to be unexpectedly rich.  It  includes a
$6$-dimensional de Sitter vacuum ($dS_6$), a variety of metastable
$4$-dimensional de Sitter and anti-de Sitter vacua ($dS_4$ and
$AdS_4$) with two extra dimensions compactified on a $2$-sphere
($S_2$), and a number of $AdS_2$ vacua with four extra dimensions
compactified on $S_4$. The model also includes some perturbatively
unstable $dS_4 \times S_2$ and $dS_2 \times S_4$ vacua (see also \cite{Krishnan:2005su}).   

Quantum tunneling transitions between all these vacua have been
discussed in
Refs. \cite{BlancoPillado:2009di,Carroll:2009dn,BlancoPillado:2009mi}.
Transitions between different $dS_4\times S_2$ vacua (which were
called ``flux tunneling transitions'' in \cite{BlancoPillado:2009di})
are similar to bubble nucleation in $4d$.  They can be studied using
the effective $4d$ theory, where they are described by the usual
Coleman-De Luccia formalism.  From a higher-dimensional viewpoint, the
compactified $S_2$ in the new vacuum inside the bubble has a different
radius and a different magnetic flux.  The role of the bubble walls is
played by magnetically charged black $2$-branes.  

\begin{figure}
\centering\leavevmode
\epsfig{file=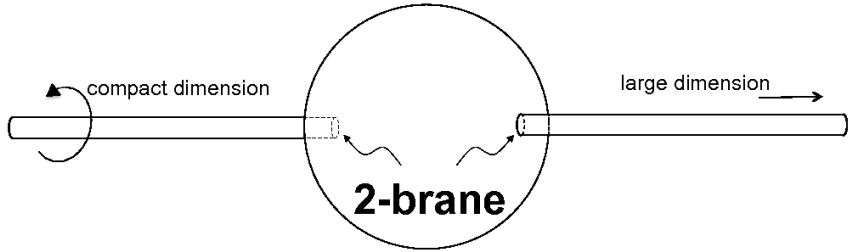,width=5cm,angle=-90}
\caption {A spatial slice through the decompactification bubble spacetime. The points marked as ``2-brane'' in this lower-dimensional analogue become a $2$-sphere in $5d$.}
\label{Alexfig5}
\end{figure}

Another kind of transition occurs when the extra dimensions are
destabilized in the new vacuum, so they start growing inside the
bubble.  The geometry resulting from such a 
decompactification transition, $dS_4 \times S_2 \rightarrow dS_6$, is
illustrated in Fig.~\ref{Alexfig5}.  Observers in the parent vacuum
see the nucleation and subsequent expansion of a bubble bounded by a
spherical black $2$-brane. The bubble interior is initially highly
anisotropic, but as the compact dimensions expand, it approaches local
isotropy, with the metric approaching that of a $6d$ de Sitter space. 

Transdimensional compactification transitions in
which the parent vacuum has higher dimensionality than the daughter
vacuum lead to a rather different spacetime structure 
\cite{Carroll:2009dn,BlancoPillado:2009mi}. 
Transitions from $dS_6$ to $dS_4\times S_2$ occur through spontaneous
nucleation of horizon-size spherical black branes, with the new
compactified regions contained in the brane interiors.  In
$dS_6\rightarrow AdS_2 \times S_4$ transitions, the new regions are
located inside spontaneously nucleated pairs of electrically charged
black holes.

Transitions from $dS_4\times S_2$ to $AdS_2\times S_4$ may also occur,
but the corresponding instantons have not yet been identified.  An
additional vacuum decay channel, which can exist in more complicated
models, is the nucleation of the so-called ``bubbles of nothing''
\cite{Witten,I-Sheng,Jose} -- that is, holes in spacetime bounded by
higher-dimensional topological defects.  In the case of flux
compactification in $6d$, such bubbles can be formed if the Maxwell
field is embedded into a gauge theory with a spontaneously broken
non-Abelian symmetry \cite{Jose}.  The boundary defect then has
codimension 3 and its cross-section has the form of a non-Abelian
magnetic monopole.


\subsection{Scale factor and volume factor times}

The scale factor and volume factor times can be defined as before, in
terms of a congruence of geodesics orthogonal to some initial
hypersurface $\Sigma_0$.  The geodesics describe a dust of test
particles which are initially uniformly spread over $\Sigma_0$.  (If
$\Sigma_0$ has some compact dimensions, the particles are assumed to
be uniformly spread both in large and compact dimensions.)  The scale
factor time is defined as 
\beq
d\eta = H(\tau) d\tau,
\eeq
where $\tau$ is the proper time along geodesics, $H$ is given by
\beq
H={1\over{D}}u^\mu_{;\mu} ,
\eeq
and $D$ is the effective number of spatial dimensions (that is, the number of
non-compact dimensions) in the corresponding vacuum. 

The volume factor time $T$ is related to $\eta$ as
\beq 
dT = Dd\eta.
\eeq
As the geodesics flow into bubbles where some dimensions get
decompactified or into black branes which have compactified regions in
their interiors, the value of $D$ changes and with it the relation
between $T$ and $\eta$.  Hence, in a transdimensional multiverse,
equal scale factor hypersurfaces are not the same as equal volume
factor hypersurfaces.

\subsection{The rate equation}

\subsubsection{Scale factor time}

Let us first consider the comoving volume rate equation using the
scale factor time $\eta$ as our time variable.  The form
of the rate equation is the same as in the equidimensional case,
\beq
{df_{ij}\over{d\eta}}= -\kappa_{i}f_{ij} + \kappa_{ij}f_j.
\label{transdfdt}
\eeq 
Here, $f_{ij}$ is the comoving volume fraction in regions of type $ij$ (which is
proportional to the number of the geodesic ``dust'' particles in such
regions) and $\kappa_{ij}$ is the probability per unit scale factor time 
for a geodesic which is currently in vacuum $j$ to cross
into vacuum $i$.  The effect of bubble or black brane nucleation is
to remove some comoving volume from the parent vacuum and add it to
the daughter vacuum (except for the bubbles of nothing, where the
excised comoving volume is lost).  
As in the $4d$ case, the asymptotic radius of black branes and bubbles 
is equal to the comoving horizon size.  This asymptotic value is
approached very quickly, essentially in one Hubble time, and since the
rate equation should be understood in the sense of coarse-graining
over a Hubble scale, we can say it is reached instantly.
The transition rates $\kappa_{ij}$ are given by
\beq
\kappa_{ij} = v_j H_j^{-(D_j +1)} \Gamma_{ij} ,
\label{transkappa}
\eeq
where $D_j$ is the effective number of dimensions in the parent vacuum
and $v_j$ is a geometric factor.  For flux tunneling
bubbles, decompactification bubbles, and bubbles of nothing, the
entire horizon region is excised, and $v_j$ is the volume
inside a unit sphere in $D_j$ dimensions,\footnote{A similar
  transdimensional rate equation, but with somewhat different 
  transition coefficients was presented by Carroll {\it
    et.al} in  Ref.~\cite{Carroll:2009dn}. In particular, they made no
  distinction between the geometric factors for compactification and
  decompactification transitions.} 
\beq
v_j = {2\pi^{(D_j)/2}\over{\Gamma\left({{D_j +3}\over{2}}
    \right)}}. 
\eeq
We note that $\Gamma_{ij}$ is the bubble (or black brane) nucleation
rate in the large $(D_j +1)$ dimensions, so that $\kappa_{ij}$ is
dimensionless. 

In the case of compactification transitions, the geometric factor is
different.  Consider, for example, transitions $dS_6 \rightarrow AdS_2
\times S_4$ in our $6d$ Einstein-Maxwell model.  These transitions
occur through nucleation of electrically charged black holes, which
are described by 6-dimensional Reissner-Nordstrom-de Sitter solutions, 
\beq
ds^2=-f(r)dt^2 +f^{-1}(r)dr^2 +r^2 d\Omega_4^2 ,
\label{RNDS}
\eeq
where
\beq
f(r) = 1-{{2{M}}\over{r^3}} + {{Q}^2\over{r^6}} - H_6^2r^2 .
\eeq
Here, $H_6$ is the expansion rate of $dS_6$ and the parameters $M$ and
$Q$ are proportional to the black hole mass and charge, respectively.
The requirement that the instanton describing black hole nucleation
should be non-singular imposes a relation between $M$, $Q$ and $H_6$.
In the limit when the black hole is much smaller than the cosmological
horizon,
\beq
r_{bh}\ll H_6^{-1} ,
\label{smallbh}
\eeq
it can be shown \cite{BlancoPillado:2009mi} that the black holes are
nearly extreme, $M\approx Q$, with a Schwarzschild radius $r_{bh}
\approx Q^{1/3}$.

Dust particles initially at rest in the metric (\ref{RNDS}) exhibit two types of
behavior.  For sufficiently small values of $r<r_*$, particles are
captured by the black hole, while for $r>r_*$ they are driven away
from the black hole by the cosmological expansion.  The critical value
$r_*$ can be found from
\beq
f'(r_*)=0.
\label{r*}
\eeq
In the limit (\ref{smallbh}), we have
\beq
r_* \approx (3Q/H_6^2)^{1/5} .
\eeq
With $D_j =5$, the geometric factor is
\beq
v_j\approx {8\over{15}}\pi^2 (H_j r_*)^6.
\eeq
This expression is not exact, since comoving geodesics in the parent
vacuum will not be at rest in the coordinates (\ref{RNDS}), but it
will give a good approximation when (\ref{smallbh}) is satisfied.

In the case of $dS_6\rightarrow dS_4\times S_2$ transitions, the
metric of nucleating black branes is
\beq
ds^2=B^2(r)[-dt^2 +\cosh^2 t~ d{\Omega'}_2^2]+dr^2
  +\rho^2(r)d\Omega_2^2. 
\eeq
The functions $B(r)$ and $\rho(r)$ are not known analytically, but
have been found numerically in Ref.~\cite{BlancoPillado:2009mi}.  Once
again, the critical radius $r_*$ below which the geodesics are
captured by the brane can be found from the condition
\beq
B'(r_*)=0,
\eeq
and in the limit of $r_* H_j\ll 1$ we have
\beq 
v_j\approx (4\pi)(4\pi r_*^3H_j^3/3).
\eeq
Here, the first factor comes from the area of the horizon-size 2-brane, and
the second factor is the fraction of the horizon volume occupied by
the captured particles in the three dimensions transverse to the brane\footnote{ We note that it follows from the analysis in Ref.'s~\cite{Carroll:2009dn,BlancoPillado:2009mi}  that in order to have a positive-energy vacuum inside the nucleated black brane, the magnetic charge of the brane should be of the order $Q_m \sim H_6^{-1}$. This implies that the black brane and cosmological horizons are comparable to one another, and therefore $r_* \sim H_6^{-1}$.  Hence, in this case the geometric factor is $v_j \sim 1$.  }.

One complication of the transdimensional case is that the volumes $V_{ij}$
occupied by different regions are no longer simply related to the
comoving volume fraction $f_{ij}$.  
The reason is that a given geodesic generally crosses regions
with different values of $D_j$ and is thus exposed to different
expansion rates at different times in its history.  The rate equation
for the physical 
volumes $V_{ij}$ on the constant-$\eta$ surfaces can be written as
\footnote{Eq.~(\ref{transsfdfdt}) assumes that transitions between
different vacua occur with conservation of volume, so the total volume
is changed only due to cosmological expansion and due to volume loss
to terminal bubbles.  This would be so for a continuous evolution of
the geodesic congruence, but is not obviously true for quantum
transitions between the vacua.  
This issue requires further study.  Here we note that accounting for a
possible discontinuous volume change in the course of transitions
could introduce an additional factor multiplying $\kappa_{ij}$ in
(\ref{transsfdfdt}).  This would result in a renormalization of
$\kappa_{ij}$ and would not change our general conclusions.}   
 \beq
{dV_{ij}\over{d\eta}}= -\kappa_{i}V_{ij} + \kappa_{ij}V_j +D_i V_{ij},
\label{transsfdfdt}
\eeq 

where we remind the reader that $\kappa_{i}=\sum_k \kappa_{ki}$ is the total decay rate out of vacuum $i$, $V_j= \sum_{k} V_{jk}$ is the physical volume fraction of type $j$ vacua, reached from all possible type $k$ parent vacua, and $D_i$ is simply the number of large inflating directions in vacuum $i$.
Here, the volumes $V_{ij}$ should be understood in the
higher-dimensional sense, that is, including both large and compact
dimensions.  With the ansatz
\beq
V_{ij} = V_0 s_{ij} e^{\gamma\eta} ,
\eeq
this can be rewritten as
\beq
(\gamma-D_i+\kappa_i)s_{ij} = \kappa_{ij}s_j ,
\label{gammakappa}
\eeq
where 
\beq
s_j = \sum_k s_{jk} .
\eeq

The right-hand side of
(\ref{gammakappa}) must be positive, since $s_j>0$ for inflating vacua. If
$i$ is also an inflating vacuum, we should have $s_{ij}>0$ and
\beq
D_i-\kappa_i <\gamma.
\label{Dkappagamma}
\eeq
Now, if we define
\beq
\gamma\equiv D_{max}-q ,
\label{gammaDmax}
\eeq
where $D_{max}$ is the largest dimension of an inflating vacuum in the
landscape, then

\beq
q< D_{max}-D_i+\kappa_i.
\label{qbound}
\eeq

The bound in Eq. (\ref{qbound}) must be satisfied for all type $i$ vacua.  The tightest constraint occurs when $D_i=D_{max}$, from which it follows that\footnote{This
  derivation of Eq.~(\ref{kappaminDmax}) follows closely the
  derivation of Eq.~(\ref{qkappamin}) in Ref.~\cite{BB1}.}
\beq
q<\kappa_{min}(D_{max}) ,
\label{kappaminDmax}
\eeq
where $\kappa_{min}(D_{max})$ is the smallest decay rate among the
vacua of effective dimension $D_{max}$.  As before, we expect
$\kappa_{min}(D_{max})$ to be very small, and thus 
\beq
\gamma\approx D_{max}.
\eeq

\subsubsection{Volume factor time}

The volume-factor-time rate equation can be written as
\beq
{df_{ij}\over{dT}}= -{\tilde\kappa}_{i}f_{ij} + {\tilde\kappa}_{ij}f_j ,
\label{transvoldfdt}
\eeq 
where the transition rates per unit time $T$ are given by
\beq
{\tilde\kappa}_{ij}={\kappa_{ij}\over{D_j}} .
\eeq

The asymptotic solution of this equation for non-terminal bubbles $ij$ is
\beq
f_{ij} \approx {\tilde s}_{ij} e^{-{\tilde q}T} ,
\label{ftildes}
\eeq
where $-{\tilde q}$ is the smallest (by magnitude) eigenvalue of the
matrix
\beq
{\tilde M}_{ij} = {\tilde\kappa}_{ij} - \delta_{ij}{\tilde\kappa}_i .
\eeq
Substitution of (\ref{ftildes}) into (\ref{transvoldfdt}) gives the
relation  
\beq
({\tilde\kappa}_i-{\tilde q}){\tilde s}_{ij} = {\tilde\kappa}_{ij}
    {\tilde s}_j ,
\label{tildekappatildes}
\eeq
and it follows, since the right hand side of (\ref{tildekappatildes}) must be positive, that ${\tilde q}$ is smaller than the
decay rate per unit volume factor time $T$ of any inflating vacuum, 
\beq
{\tilde q} < {\tilde\kappa}_{min} .
\eeq

Unlike the scale factor case, the volumes occupied by different vacua
on the constant-$T$ surfaces are simply related to the comoving volume
fractions,
\beq
V_{ij}(T)=V_0 f_{ij}(T) e^T = V_0 {\tilde s}_{ij} e^{{\tilde\gamma}T},
\eeq
where
\beq
{\tilde\gamma}=1-{\tilde q}\approx 1.
\label{gammatildeq}
\eeq

\subsection{The number of observers}

\subsubsection{Scale factor cutoff}

The number of observers can now be evaluated along the same lines as
in Section IIB.  For the scale factor cutoff, the rate of volume increase in type-$ij$ bubbles due to new bubble formation is
\beq
d^+ V_{ij}=V_0 \kappa_{ij}s_j e^{\gamma\eta}d\eta ,
\eeq
and the total volume of ``observation hypersurfaces" in $ij$ bubbles is given by
\begin{eqnarray}
V_{ij}^{obs} = V_0 e^{D_i\Delta\eta_{ij}^{obs}}
e^{-\lambda_{ij}\Delta\eta_{ij}^{obs}}
\int_0^{\eta_{max}-\Delta\eta_{ij}^{obs}} \kappa_{ij} s_j
e^{\gamma\eta} d\eta \\
= {1\over{\gamma}}V_0 \kappa_{ij}s_{j} e^{\gamma\eta_{max}}
e^{(D_i-\gamma-\lambda_{ij})\Delta\eta_{ij}^{obs}} .
\end{eqnarray}
Dropping irrelevant constant factors, multiplying by the density of
observers $n_{ij}^{obs}$ and using Eq.~(\ref{gammaDmax}), we obtain
for the total number of observers 
\beq
\mathcal{N}_{ij}^{obs}\propto \kappa_{ij}s_{j} n_{ij}^{obs} 
e^{(D_i-D_{max}+q-\lambda_{ij})\Delta\eta_{ij}^{obs}} .
\label{Nobssfc}
\eeq
Note that here the density $n_{ij}^{obs}$ should be understood in a
higher-dimensional sense, so that $n_{ij}^{obs}\propto 1/V_i^{comp}$,
where $V_i^{comp}$ is the volume of compact dimensions in vacuum $i$.

\subsubsection{Volume factor cutoff}

Similarly, for the volume factor cutoff we find
\beq
d^+ V_{ij}=V_0 {\tilde\kappa}_{ij}{\tilde s}_j e^{{\tilde\gamma}T}dT ,
\eeq
\begin{eqnarray}
V_{ij}^{obs} = V_0 e^{\Delta T_{ij}^{obs}}
e^{-{\tilde\lambda}_{ij}\Delta T_{ij}^{obs}}
\int_0^{T_{max}-\Delta T_{ij}^{obs}} {\tilde\kappa}_{ij} {\tilde s}_j
e^{{\tilde\gamma}T} dT \\
= {1\over{\tilde\gamma}}V_0 {\tilde\kappa}_{ij}{\tilde s}_{j} 
e^{{\tilde\gamma}T_{max}}
e^{({\tilde q}-{\tilde\lambda}_{ij})\Delta T_{ij}^{obs}} ,
\end{eqnarray}
and finally
\beq
\mathcal{N}_{ij}^{obs}\propto {\tilde\kappa}_{ij}{\tilde s}_{j} 
n_{ij}^{obs} e^{({\tilde q}-{\tilde\lambda}_{ij})\Delta T_{ij}^{obs}} .
\label{Nobsvfc}
\eeq
Here,
\beq
{\tilde\lambda}_{ij} = \lambda_{ij}/D_i
\eeq
and
\beq
\Delta T_{ij}^{obs} \approx D_i \Delta \eta_{ij}^{obs}
\label{Tij}
\eeq
is the volume factor time it takes for observers to evolve in
type-$ij$ bubbles.

\section{Properties of scale factor and volume factor measures}

\subsection{Number of e-folds of inflation and the density parameter
  $\Omega$} 

Suppose we have a landscape which includes bubbles (or black branes)
with comparable nucleation rates and similar post-inflationary
evolution, but different inflaton potentials.  The bubbles then have
different numbers ${\cal I}$ of e-folds of slow-roll inflation
and different values for the density parameter $\Omega$ (measured at
the same value of the CMB temperature, $T_{CMB}=2.7~K$).\footnote{The
  primordial density contrast $Q$ will also generally vary.  To
  simplify the analysis, here we consider only the section of the
  landscape where $Q$ is fixed to its observed value in our part of
  the universe.  We have verified, using some simple models, that
  inclusion of variation of $Q$ does
  not significantly alter our conclusions.}   
We want to investigate the effect of these differences on the number
of observers, ${\cal N}$, regulated with scale factor and volume
factor cutoffs.   
 
The scale factor time $\Delta\eta_{ij}^{obs}$, which appears in
Eq.~(\ref{Nobssfc}) for the number of observers, can be expressed as 
\beq
\Delta\eta_{ij}^{obs} \approx {\cal I}_{ij} + \Delta{\eta}_{ij}' ,
\label{Iij}
\eeq
where $\Delta\eta_{ij}'$ is the time it takes for observers to evolve
after the end of inflation.  The dependence of ${\cal N}_{ij}$ on the
number of inflationary e-folds ${\cal I}_{ij}$ can now be read off
from Eqs.~(\ref{Nobssfc}), (\ref{Iij}), 
\beq
\mathcal{N}_{ij}^{obs}\propto 
e^{(D_i-D_{max}+q-\lambda_{ij}){\cal I}_{ij}} .
\label{NijD}
\eeq

For an equidimensional landscape with $D_i = D_{max} = 3$,
Eq.(\ref{NijD}) gives
\beq
{\cal N}_{ij}^{obs} \propto 
e^{(q-\lambda_{ij}){\cal I}_{ij}} \approx 1 .
\label{NI}
\eeq
Since $q$ and $\lambda_{ij}$ are both extremely small, there is no
selection effect favoring small or large values of ${\cal I}_{ij}$
\cite{DeSimone:2008bq,DeSimone:2009dq}.  
What has happened here is that there has been an almost precise
cancellation of two effects: the volume growth inside bubbles, and the
decrease in the number of nucleated bubbles as we go back in
time.

Apart from the explicit dependence on ${\cal I}_{ij}$ shown in Eq.
(\ref{NijD}), $\mathcal{N}_{ij}^{obs}$ also depends on ${\cal I}_{ij}$
through the anthropic factor $n_{ij}^{obs}$ and through the ``prior"
distribution  
\beq
{\cal P}_{prior}\propto \kappa_{ij}s_j .  
\eeq
Since a long slow-roll inflation requires fine-tuning, one can expect
that ${\cal P}_{prior}({\cal I})$ is a decreasing function of ${\cal
  I}$.  Analysis of a simple landscape model in Ref.~\cite{FKRS}
suggests a power-law dependence, 
\beq
d{\cal P}_{prior}\propto {\cal I}^{-\alpha}d{\cal I} 
\label{prior}
\eeq
with $\alpha = 4$.

The dependence of the anthropic factor $n^{obs}$ on ${\cal I}$ has
been  discussed in Refs.~\cite{GTV,FKRS,DeSimone:2009dq}.  This dependence is
more conveniently expressed in terms of the density parameter
$\Omega$, which is related to ${\cal I}$ as 
\beq
1-\Omega = e^{2({\cal C} - {\cal I})}
\label{OmegaI}
\eeq
The value of the constant ${\cal C}$ depends on the duration of the
inflaton-dominated period, which may occur between the end of inflation and the onset of thermalization, and on the
thermalization temperature $T_{th}$.  Assuming instant thermalization
at a GUT-scale temperature $T_{th}\sim 10^{15}~GeV$, we have ${\cal
  C}\approx 61.5$.  In the opposite extreme of low thermalization
temperature, $T_{th}\sim 10^3~GeV$, ${\cal C}\approx 34$. 

The anthropic significance of $\Omega$ is due to its effect on
structure formation.  For $\Omega$ very close to 1, $(1-\Omega)\ll 1$,
this effect is negligible.  For $\Omega$ significantly less than 1,
the growth of density perturbations terminates at redshift
$(1+z_{\Omega})\approx \Omega^{-1}$.  For $z_\Omega \gtrsim 1$ this
strongly interferes with structure formation.  The
density of galaxies, and the density of observers, decreases somewhat
with $\Omega$ in the range $0.9 \gtrsim \Omega \gtrsim 0.4$ and drops
rapidly towards zero at $\Omega \lesssim  \Omega_c \sim 0.3$
\cite{DeSimone:2009dq}. 
This behavior can be roughly accounted for by imposing a sharp cutoff at
$\Omega_c\approx 0.3$.  Using Eq.~(\ref{OmegaI}), this translates into
a cutoff on ${\cal I}$ at ${\cal I}\approx  {\cal C} -
{1\over{2}}\ln(0.7) \approx {\cal C}$.  The density
of observers then effectively becomes a step function,
\beq
n^{obs}\propto \theta({\cal I}-{\cal C}) .
\label{step}
\eeq
With the prior from Eq.(\ref{prior}), the full distribution for ${\cal
  I}$ is given by 
\beq
d{\cal P}({\cal I}) =
(\alpha -1){\cal C}^{\alpha -1} \theta({\cal
  I}-{\cal C}) {\cal I}^{-\alpha}d{\cal I} ,
\label{PI}
\eeq
where the prefactor is chosen so that the distribution is normalized
to 1.  

Observations indicate that in our part of the universe
$|1-\Omega|\lesssim 0.01$, 
which corresponds to 
\beq
{\cal I}> {\cal I}_0 = {\cal C}+\ln 10 
\label{I0}
\eeq
According to (\ref{PI}), the probability of this observation is
\beq
{\cal P}({\cal I}> {\cal I}_0) = ({\cal C}/{\cal I}_0)^{\alpha -1} .
\eeq
Since ${\cal I}_0$ in Eq.~(\ref{I0}) is only slightly greater than
${\cal C}$, this probability is close to 1, as long as $\alpha$ is not
very large (which is certainly the case for $\alpha=4$).  
Hence, the observed value
of $\Omega$ is fully consistent with the equidimensional scale factor
measure \cite{DeSimone:2009dq}.   
  
The situation is very different in a transdimensional landscape.  
Omitting the exponentially small terms in the exponent of
Eq.~(\ref{NijD}), we have  
\beq
{\cal N}_{ij}^{obs}\propto e^{-(D_{max}-D_i){\cal I}_{ij}} .
\label{NDI}
\eeq
This shows that large amounts of slow-roll inflation are strongly
disfavored in all bubbles with $D_i < D_{max}$.  For example, in the
case of $(3+1)$-dimensional bubbles or branes $(D_i =3)$ in our $6d$
Einstein-Maxwell theory $(D_{max}=5)$, ${\cal N}_{ij}^{obs}\propto
e^{-2{\cal I}_{ij}}$. 
In string theory landscape, we could have $D_{max}=9$ and an even
stronger suppression, ${\cal N}_{ij}^{obs}\propto e^{-6{\cal I}_{ij}}$. 

The reason is as follows:  bubbles which undergo less e-folds of
inflation before being able to produce observers get rewarded by being
more plentiful, but undergo less expansion than other bubbles which
have a longer ``gestation" period.  These are the same two competing
effects we discussed above for regular $(3+1)d$ inflation.  The
difference here, in the transdimensional case, is that the internal
volume reward (which grows like $e^{D_i {\cal I}_{ij}}$) is not enough
to compensate for the lower number of bubbles which nucleate at earlier
times  ($\propto e^{-D_{max}{\cal I}_{ij}}$).   

Using Eq.~(\ref{step}) for $n^{obs}$ and disregarding the power-law
prior compared to the steep dependence on ${\cal I}$ in (\ref{NDI}),
we obtain
\beq
d{\cal P}=(D_{max}-D) \theta({\cal I}-{\cal C}) 
e^{-(D_{max}-D)({\cal I}-{\cal C})} d{\cal I} .
\eeq
The probability for the observed bound on $\Omega$ is then
\beq
{\cal P}({\cal I}>{\cal I}_0) =e^{-(D_{max}-3)({\cal I}_0-{\cal C})} =
10^{-(D_{max}-3)} ,
\eeq
where we have set $D=3$ and used Eq.~(\ref{I0}) in the last step.
For $D_{max}> 5$, this probability is smaller than $10^{-3}$, so the
transdimensional scale factor measure can be ruled out at a high
confidence level.

Turning now to the volume factor measure, we have from
Eqs.~(\ref{Iij}),(\ref{Tij}) 
\beq
\Delta T_{ij}^{obs} \approx D_i({\cal I}_{ij}+\Delta\eta_{ij}') ,
\eeq
and Eq.~(\ref{Nobsvfc}) for the number of observers gives
\beq
\mathcal{N}_{ij}^{obs}\propto
e^{({\tilde q}-{\tilde\lambda}_{ij})D_i{\cal I}_{ij}} \approx 1 .
\eeq
This is essentially the same as Eq.(\ref{NI}).  There is no strong selection
for large or small values of ${\cal I}$, so the same analysis as we
gave above for the equidimensional case should apply here.  Thus, we
conclude that the transdimensional volume factor measure is consistent
with observations of $\Omega$.

\subsection{Youngness bias}

Another pitfall that a candidate measure needs to avoid is the
``youngness paradox'' \cite{LLM95,Guth00}.  Consider two kinds of
observers who live in the same type $ij$ of bubbles, but take
different time $\Delta\eta_{ij}'$ to evolve after the end of
inflation.  Observers who take less time to evolve can live in bubbles
that nucleate closer to the cutoff and thus have more volume available
to them.  With proper time slicing, this growth of volume is so fast
that observers evolving even a tiny bit faster are rewarded by a huge
volume factor, resulting in some bizarre predictions
\cite{LLM95,Guth00,Tegmark,Bousso:2007nd}.  On the other hand, the scale factor
cutoff in $(3+1)d$ gives only a mild youngness bias \cite{DeSimone:2008bq},
\beq
\mathcal{N}_{ij}^{obs}\propto 
e^{-(3-q+\lambda_{ij})\Delta\eta_{ij}'} \approx e^{-3\Delta\eta_{ij}'} ,
\label{youngness3d}
\eeq
so observers evolving faster by a scale factor time $\delta\eta$ are
rewarded by a bias factor
\beq
B\approx e^{3\delta\eta} .
\label{bias3d}
\eeq

The generalization of Eqs.~(\ref{youngness3d}) and (\ref{bias3d}) to a
transdimensional landscape can be read off from Eq.~(\ref{Nobssfc}):
\beq
\mathcal{N}_{ij}^{obs}\propto 
e^{-(D_{max}-q+\lambda_{ij})\Delta\eta_{ij}'} \approx e^{-D_{max}\Delta\eta_{ij}'} ,
\label{youngness}
\eeq
and
\beq
B\approx e^{D_{max}\delta\eta} .
\label{bias}
\eeq 
Note that here, and in Eq.(\ref{youngness3d}) above, 
 we omitted the factor $\exp(D_i\Delta\eta_{ij}')$
which accounts for the growth of volume inside bubbles in the time
interval $\Delta\eta_{ij}'$.  The reason is that post-inflationary
expansion only dilutes matter (e.g., galaxies) and does not by itself
result in any increase in the number of observers.\footnote{This means
  that the density of observers that we used in the preceding sections
  should be $n_{ij}^{obs}\propto \exp(-D_i\Delta\eta_{ij}')$.  This
  relation accounts only for dilution of galaxies with cosmic time,
  but in fact some new galaxies will also be formed, so the dependence
  of $n_{ij}^{obs}$ on $\Delta\eta_{ij}'$ will be more complicated.
  This effect is unimportant in the context of the youngness paradox,
  so we disregard it here.}
In string theory landscape we could have $D_{max}=9$, so the youngness
bias (\ref{bias}) is stronger than in the equidimensional $(3+1) d$ case, but it
is still mild enough to avoid any paradoxes.

Similarly, for the transdimensional volume factor cutoff we have
\beq
\mathcal{N}_{ij}^{obs}\propto 
e^{-(1-{\tilde q}+{\tilde\lambda}_{ij})D_i\Delta\eta_{ij}'} \approx
e^{-D_i\Delta\eta_{ij}'} 
\eeq
and
\beq
B\approx e^{D_i\delta\eta} .
\label{biasD}
\eeq 
For $(3+1)$-dimensional regions this gives $B\approx
\exp(3\delta\eta)$, the same as in the equidimensional case.

\subsection{Boltzmann brains}

Yet another potential problem for the measure is an excessive
production of Boltzmann brains (BBs) -- freak observers spontaneously
nucleating as de Sitter vacuum fluctuations
\cite{Rees1,DKS02,Albrecht,Page1,Page06,BF06}.  
Measures predicting that BBs greatly outnumber ordinary
observers should be ruled out.

The total number of BBs nucleated in type-$ij$ bubbles can be
estimated as
\beq
\mathcal{N}_{ij}^{BB}= \Gamma_i^{BB} \Omega_{ij} ,
\label{NBB}
\eeq
where $\Omega_{ij}$ is the total spacetime volume in such bubbles and
$\Gamma_i^{BB}$ is the BB 
nucleation rate.  Here we disregard possible time variation of the
nucleation rate during the early evolution in bubble interiors.  The
bulk of BB production tends to occur at late stages, after vacuum
domination.  The nucleation rate is then constant and depends only on
the vacuum inside the bubble, not on the parent vacuum.

The spacetime volume $\Omega_{ij}$ can be found by integrating the
spatial volume over the proper time $\tau$.  With a scale factor
cutoff, we find  
\beq
\Omega_{ij} = \int V_{ij} d\tau = H_i^{-1}\int^{\eta_{max}}V_{ij}(\eta)d\eta
= {V_0\over{\gamma H_i}}s_{ij} e^{\gamma\eta_{max}} ,
\label{Oij}
\eeq
with $\gamma$ from Eq.~(\ref{gammaDmax}).  Combining this with
Eqs.~(\ref{NBB}) and (\ref{Nobssfc}), we obtain for the relative number of
BBs and normal observers
\begin{eqnarray}
{\mathcal{N}_{ij}^{BB} \over{\mathcal{N}_{ij}^{obs}}} \approx
{\Gamma_i^{BB}\over{H_i}}{s_{ij}\over{\kappa_{ij}s_j n_{ij}^{obs}}} 
e^{(D_{max}-D_i)\Delta\eta_{ij}^{obs}} \\ 
\sim {\kappa_i^{BB} H_i^3 e^{(D_{max}-D_i)\Delta\eta_{ij}^{obs}} \over{
    (D_{max} -D_i -q +\kappa_i) n_{ij}^{obs}}} ,
\label{NBBNobs}
\end{eqnarray}
where 
\beq
\kappa_i^{BB}=\Gamma_i^{BB}{4\pi\over{3}}H_i^{-4}
\eeq
is the dimensionless BB nucleation rate and we have used
Eq.~(\ref{gammakappa}) in the last step. 

In an equidimensional landscape with $D_i=D_{max}$, Eq.~(\ref{NBBNobs}) gives
\beq
{\mathcal{N}_{ij}^{BB} \over{\mathcal{N}_{ij}^{obs}}} \sim
{\kappa_i^{BB} \over{\kappa_i}}{H_i^3\over{n_{ij}^{obs}}} 
\eeq
The rate of BB nucleation $\kappa_i^{BB}$ is extremely small, with
estimates ranging from $\exp(-10^{16})$ to $\exp(-10^{120})$
\cite{BB1,BB2}.  Compared to this, most quantities which can be large
or small by ordinary standards, can be regarded as $O(1)$.  In
particular, this applies to the number of observers per horizon
volume, $n_{ij}^{obs}H_i^{-3}$ \cite{BB1,BB2}.  Hence,  
\beq
{\mathcal{N}_{ij}^{BB} \over{\mathcal{N}_{ij}^{obs}}} \sim
{\kappa_i^{BB} \over{\kappa_i}} ,
\label{kappaBBkappa}
\eeq
so BB domination in vacuum $i$ is avoided only if the vacuum decay
rate is greater than the BB nucleation rate, 
\beq
\kappa_i \gtrsim \kappa_i^{BB} .
\label{BBcondition}
\eeq
In particular, this condition should be satisfied by our vacuum \cite{Linde06}.

Even if ordinary observers can evolve only in our vacuum,  BBs might
be able to nucleate in other vacua as well.  For example, vacua having
the same low-energy physics and differ from ours only by the value of
the cosmological constant $\Lambda$ can nucleate human-like observers,
as long as $\Lambda$ is small enough, so that the observers are not
torn apart by the gravitational tidal forces.   
General conditions for the avoidance of BB domination have been
discussed in Refs.~\cite{BB1,BB2}.  The main condition is that
Eq.~(\ref{BBcondition}) should be satisfied for any vacuum $i$ that
can support Boltzmann brains.  This is a rather non-trivial condition.
Vacuum decay rates can be extremely low, especially for vacua with a
low supersymmetry breaking scale.  Hence, the equidimensional scale
factor measure may or may not be consistent with observations,
depending on the details of the landscape. 

In the transdimensional case, for vacua with $D_i<D_{max}$,
Eq.~(\ref{NBBNobs}) gives 
\beq
{\mathcal{N}_{ij}^{BB} \over{\mathcal{N}_{ij}^{obs}}} 
\sim \kappa_i^{BB} \ll 1 ,
\eeq
where, following \cite{BB1,BB2}, we have set
$e^{(D_{max}-D_i)\Delta\eta_{ij}^{obs}}\sim 1$. 
Thus, the Boltzmann brain problem of the scale factor measure
disappears in the transdimensional case, even if it existed in the
equidimensional landscape. 

For the volume factor cutoff, Eqs.~(\ref{Oij}) and (\ref{NBBNobs}) are
replaced by  
\beq
\Omega_{ij} = {1\over{D_i H_i}}\int^{T_{max}}V_{ij}(T)dT
= {V_0\over{{\tilde\gamma}D_i H_i}}{\tilde s}_{ij} e^{{\tilde\gamma}T_{max}} ,
\label{OijT}
\eeq
and
\beq
{\mathcal{N}_{ij}^{BB} \over{\mathcal{N}_{ij}^{obs}}} \approx
{\kappa_i^{BB}H_i^3\over{D_i n_{ij}^{obs}}} 
{e^{({\tilde\lambda}_{ij}-{\tilde q})\Delta T_{ij}^{obs}}
\over{{\tilde\kappa}_{i}-{\tilde q}}} 
\sim {\kappa_i^{BB} \over{\tilde\kappa}_i} .
\eeq
This is essentially the same as Eq.~(\ref{kappaBBkappa}) for the
equidimensional scale factor measure.  Hence, we expect the conditions
for BB avoidance in a transdimensional landscape with a volume factor
cutoff to be the same as derived in \cite{BB1,BB2} for the scale
factor measure in $(3+1)d$.

\section{Conclusions}

In this paper we investigated possible generalizations of the scale
factor measure to a transdimensional multiverse.  We considered two
such generalizations: a straightforward extension of the scale factor
cutoff and the volume factor cutoff.  Here we briefly summarize and
discuss our results.

We found that the transdimensional scale factor measure is not
subject to the youngness or Boltzmann brain paradoxes.  
One of the key features of this measure is that
it exponentially disfavors large amounts of slow-roll inflation inside
the bubbles. This results in preference for low values for the
density parameter $\Omega$, while observations indicate that
$\Omega$ is actually very close to 1, 
\beq 
|1-\Omega_0|\lesssim 0.01 .
\label{Omegabound}
\eeq

The severity of the problem depends on the highest dimension $D_{max}$
of an inflating vacuum in the landscape.  The probability for the
bound (\ref{Omegabound}) to hold is
\beq
{\cal P}_0 \sim 10^{D_{max}-3}.
\eeq
For $D_{max}=4$ this gives a reasonably large probability ${\cal P}_0
\sim 0.1$. If indeed we live in a multiverse with $D_{max}=4$, then
this measure makes a strong prediction that a negative curvature will
be detected when the measurement accuracy is improved by another order
of magnitude.  On the other hand, if $D_{max}>5$, as can be expected
in string theory 
landscape, we have ${\cal P}_0 \lesssim 10^{-3}$, and the scale factor
measure is observationally ruled out at a high confidence level.

One potential pitfall for the measure that we have not discussed in
the body of the paper
is the runaway problem, also known as $Q$ or $G$ catastrophe.  This
problem arises in measures that exponentially favor large amounts of
inflationary expansion.  The primordial density contrast $Q$ and the
gravitational constant $G$ are strongly correlated with the number of
e-folds of inflation ${\cal I}$, and an exponential probability
distribution for ${\cal I}$ indicates that the distributions for $Q$
and $G$ should also be very steep, driving their expected magnitudes
to extreme values \cite{FHW,QGV,GS}.  These problems are often
referred to as $Q$ and $G$ ``catastrophes''.  In our case, large
inflation is exponentially ${\it disfavored}$, and the situation is
different.  The runaway toward small values of ${\cal I}$ is stopped
by the anthropic bound on $\Omega$.  The resulting prediction for
$\Omega$ may be too low, as described above.  But if the distribution
for $\Omega$ is acceptable, as it could be for $D_{max}=4$, then $Q$
and $G$ catastrophes may also be avoided.  This would happen if the
distribution is cut off by the anthropic bound on $\Omega$ before it
gets into the dangerous range of $Q$ or $G$.  (Note that $\Omega$ is
much more sensitive to the value of ${\cal I}$ than $Q$ or $G$.)

The properties of the volume factor measure are essentially the same
as those of the scale factor measure in $(3+1)$ dimensions.  It is
free from the $Q$ and $G$ catastrophes and from the youngness paradox.
It may or may not have a Boltzmann brain (BB) problem, depending on
the properties of the landscape.  The conditions for BB avoidance for
this measure are the same as found in Refs.~\cite{BB1,BB2} in the
$(3+1)d$ case, the main condition being that the rate of BB nucleation
in any vacuum should be smaller than the decay rate of that vacuum.
Whether or not this condition is satisfied in the string theory
landscape is presently an open question.  The answer depends both on
the unknown properties of the landscape and on our understanding of
what exactly constitutes a Boltzmann brain.  In any case, at our
present state of knowledge, the volume factor measure is certainly not
ruled out by the BB problem.

Our goal in this paper was to analyze and compare the properties of
the scale factor and volume factor measures in a transdimensional
multiverse.  Our results suggest that the volume factor cutoff is a
more promising measure candidate.  It avoids all potential measure
problems, with the judgement still pending on the BB paradox.  On the
other hand, we found that the scale factor measure suffers from
a serious problem, predicting low values for the density parameter
$\Omega$.  This problem becomes acceptably mild only if the highest
dimension of the inflating vacua in the landscape is $D_{max}\leq 4$.

We note finally the recent work \cite{GV08,GV09,Fiol} which used
holographic ideas for motivating the scale factor measure in $(3+1)d$.  
The proposal is that the dynamics of an eternally inflating universe
has a dual representation in the form of Euclidean field theory
defined at future infinity.  An ultraviolet cutoff in this boundary
theory then corresponds to a scale factor cutoff in the bulk.  It
would be interesting to see whether or not these ideas can be
extended to a transdimensional multiverse.

\section{Acknowledgements}

We are grateful to Jaume Garriga for discussions and some crucial
suggestions.  We also benefited from very helpful discussions with Jose
Blanco-Pillado and Michael Salem.  This work was supported in part by grant PHY-0855447
from the National Science Foundation and by grant RFP2-08-30 from the Foundational Questions Institute (fqxi.org).

\end{document}